\documentclass[12pt,letterpaper]{article}

\usepackage{amsmath,amssymb}
\usepackage{layout}
\usepackage{graphicx}

\newcommand{\rom}[1]{\mathrm{#1}}
\newcommand{\R}{\mathbb{R}}
\newcommand{\p}{\partial}
\newcommand{\al}{\alpha}
\newcommand{\ep}{\epsilon}
\newcommand{\mlrum}{\vspace{4mm}}

\newcommand{\bea}{\begin{eqnarray}}
\newcommand{\eea}{\end{eqnarray}}
\newcommand{\beastar}{\begin{eqnarray*}}
\newcommand{\eeastar}{\end{eqnarray*}}

\newcommand{\refRef} [1]{\cite{#1}}
\newcommand{\refRefs}[1]{\cite{#1}}
\newcommand{\refeq}  [1]{(\ref{#1})}
\newcommand{\refeqs} [2]{(\ref{#1}) and (\ref{#2})}
\newcommand{\refsect}[1]{section \ref{#1}}

\newcommand{\refappe}[1]{appendix \ref{#1}}
\newcommand{\reffig}[1]{figure \ref{#1}}

\begin{document}
\begin{titlepage}
\bigskip
\rightline{}
\rightline{hep-th/0305247}
\bigskip\bigskip\bigskip\bigskip
\centerline{\Large \bf {A Charged Rotating Black Ring}}
     \bigskip\bigskip
           \bigskip\bigskip

  \centerline{\large Henriette Elvang}
      \bigskip\bigskip
  \centerline{\em Department of Physics, UCSB, Santa Barbara, CA 93106}
   \centerline{elvang@physics.ucsb.edu}
                \bigskip\bigskip
\centerline{\today}
                \bigskip\bigskip
\begin{abstract}
We construct a supergravity solution describing a charged rotating
black ring with $S^2\times S^1$ horizon in a five dimensional
asymptotically flat spacetime.
In the neutral limit the solution is the rotating black ring recently
found by Emparan and Reall. 
We determine the exact value of the lower bound on $J^2/M^3$, where
$J$ is the angular momentum and $M$ the mass; the black ring
saturating this bound has maximum entropy for the given mass. 
The charged black ring is characterized by mass $M$, angular
momentum $J$, and electric charge $Q$, and it also carries local
fundamental string charge. 
The electric charge distributed uniformly along the ring helps 
support the ring against its gravitational self-attraction, so that 
$J^2/M^3$ can be made arbitrarily small while $Q/M$ remains finite. 
The charged black ring has an extremal limit in which the horizon
coincides with the singularity. 
\end{abstract}
\end{titlepage}
\baselineskip=16pt



\setcounter{equation}{0}
\section{Introduction}

Recently, Emparan and Reall \refRef{Emparan:2001wn} found an exact
vacuum solution describing a rotating black ring in a five dimensional
asymptotically flat spacetime. 
The black ring solution is the first explicit example of
non-uniqueness in higher dimensional gravity in the sense that the
asymptotically determined quantities do not uniquely specify the
solution: in five dimensions there exist asymptotically flat vacuum
solutions with the same mass and angular momentum, but with distinct
horizon topologies --- one is the rotating black hole with $S^3$
horizon and the other is the rotating black ring with $S^2 \times S^1$
horizon. We extend this non-uniqueness result to charged solutions of
low energy heterotic string theory.

In five dimensions, the Myers-Perry black hole \refRef{Myers:un}
is characterized by the mass $M$ and two
independent angular momenta, $J_1$ and $J_2$. Taking $J=J_1$ and
$J_2=0$, the dimensionless ratio constructed from $J$ and $M$ has an
upper bound,
\beastar
  \frac{J^2}{M^3} \le \frac{32}{27\pi} 
  ~~~~~~\rom{(black~hole)}\, .
\eeastar
Emparan and Reall showed that for the black ring with mass $M$ and
angular momentum $J$ the dimensionless ratio $J^2/M^3$ has a lower
bound,
\beastar
  \frac{J^2}{M^3} \ge k \, \frac{32}{27\pi} 
  ~~~~~~\rom{(neutral~black~ring)}\, .
\eeastar
and they found $k \approx 0.8437$ \refRef{Emparan:2001wn}. In this
paper we show that the exact value is \mbox{$k = 27/32$}. The
solution with $J^2/M^3 = 1/\pi$ is the black ring that
maximizes the entropy for the given mass. For 
$\frac{1}{\pi} \le \frac{J^2}{M^3} \le \frac{32}{27 \pi}$
there are spherical black hole solutions and black ring solutions with
the same values of $J$ and $M$.

The interpretation of the lower bound on $J^2/M^3$ is that for a given 
mass it takes a certain angular momentum to balance the gravitational
self-attraction of the ring. An electric charge distributed uniformly
around the black ring would help support the ring so that the ratio
$J^2/M^3$ could be made arbitrarily small. We find that this is indeed
possible. 

Applying the solution generating techniques of Hassan and Sen
\refRef{Hassan:1991mq} to the rotating black ring, we find a
solution describing a charged rotating black ring. This is a solution 
of the low energy limit of heterotic string theory (heterotic
supergravity) and besides carrying a U(1) electric charge $Q$,
the black ring also carries local fundamental string charge. 
The charged black ring can be viewed as the field of a rotating
excited loop of fundamental string with electric charged added.
We find that the ratio $J^2/M^3$ can be made arbitrarily small, while
the dimensionless ratio of charge to mass approaches a constant. 
The charge and mass satisfy $|Q| \le M$, independent of the angular
momentum. We also compute the magnetic moment and gyromagnetic ratio.
The charged rotating black ring has an extremal limit for which the
ring is extremally charged, $|Q| = M$. For the extremal solution, the
horizon coincides with the singularity.

It is unknown if supersymmetric black rings exist. In
\refRef{Gauntlett:2002nw} all supersymmetric solutions of $N=1$
minimal supergravity in five dimensions are constructed. Furthermore,
as a first uniqueness result, it is argued in \refRef{Reall:2002bh}
that the only supersymmetric, asymptotically flat black hole solutions
in this theory are the Breckenridge, Myers, Peet, and Vafa (BMPV) 
\refRef{Breckenridge:1996is} black holes, which are characterized by
their mass and angular momentum. However, we have here a black ring
solution not to minimal supergravity, but to heterotic supergravity
with five dimensions compactified. 
Hence the uniqueness result of \refRef{Reall:2002bh} does not exclude
the possibility of a supersymmetric black ring for which the
extremal limit of the black ring found in this paper may be a
candidate. 

The matter of uniqueness is interesting in its own right, but is also
important for the string theory calculations of the entropy of
supersymmetric or nearly supersymmetric black holes
\refRef{Strominger:1996sh}. For these 
derivations it is assumed that the black hole solutions are specified
uniquely by their asymptotic charges. 

The solutions presented in this paper are \emph{not} uniquely 
specified by their asymptotic charges. For a certain range of
parameters there are charged rotating black rings and spherical black
holes (obtained by applying the Hassan-Sen transformation to a
Myers-Perry black hole with a single nonzero angular momentum) with
the same asymptotic charges. The extremal limit of the charged black
ring hasvanishing horizon area.

The paper is organized as follows. We consider the neutral rotating
black ring in \refsect{s:BR}:
in \refsect{s:revBR} we review the
neutral black ring of \refRef{Emparan:2001wn} and in \refsect{s:exact}
we derive the exact lower bound of $J^2/M^3$ for the black ring.
We review the Hassan-Sen solution generating technique in
\refsect{s:review}. The Hassan-Sen technique gives the
transformed solutions implicitly. Starting from quite general
solutions we offer in \refappe{s:static} and
\refappe{s:stationary} explicit expressions for the transformed
solutions. In \refsect{s:chargedBR}, we apply the Hassan-Sen
transformation to the rotating black ring to obtain the charged ring
solution (\refsect{s:soln}). We investigate the physical properties
(\refsect{s:physquan}) and study an extremal 
limit of the charged black ring (\refsect{s:extring}). 
We compare the local behavior of the charged black ring to the local
behavior of a charged black string obtained by applying the Hassan-Sen 
transformation to a boosted black string (\refsect{s:boosted}).
Somewhat unrelated to the black rings we discuss in
\refsect{s:charBS} solutions for charged black strings
and their extremal limits. 
We summarize and discuss the results in \refsect{s:disc}.

\setcounter{equation}{0}
\section{The Black Ring}
\label{s:BR}

Emparan and Reall found vacuum solutions describing black rings with
$S^2 \times S^1$ horizons in a five dimensional asymptotically flat
spacetime. The static black ring solution  
\refRefs{Emparan:2001wk} has conical singularities preventing the ring
from collapsing, but these conical singularities can be avoided if
the ring is rotating fast enough to provide a
force to balance the ring under its own gravitational attraction 
\refRef{Emparan:2001wn}. We review the rotating black
ring in \refsect{s:revBR}, and in \refsect{s:exact} we derive the exact
lower bound on $J^2/M^3$.


\subsection{Review of the Neutral Rotating Black Ring}
\label{s:revBR}
The metric of the black ring was obtained by a Wick rotation of a
metric in \refRef{Chamblin:1996kw}. The solution is characterized by a
parameter $\nu$ and a scaling $A$. Written in C-metric coordinates 
(we adopt the notation of and follow closely
Ref.~\refRef{Emparan:2001wn}) the metric is
\bea 
 \nonumber
  ds^2 &=& -\frac{F(x)}{F(y)} 
    \left( 
      dt + \sqrt{\frac{\nu}{\xi_1}} \, \frac{\xi_2-y}{A} d\psi \,
    \right)^2 \\[2mm] \label{rotBR}
    &&\hspace{5mm}
    + \frac{1}{A^2 (x-y)^2}
    \bigg[ 
      - F(x) 
      \left( 
      G(y) d\psi^2
      + \frac{F(y)}{G(y)} dy^2
      \right) \\[2mm] \nonumber
    &&\hspace{3.3cm}
      + F(y)^2
      \left( 
      \frac{dx^2}{G(x)} 
      + \frac{G(x)}{F(x)} d\phi^2
      \right)
      \bigg] \, ,
\eea
where
\bea \label{FandG}
  F(\xi) = 1 - \xi/\xi_1 \, ,
  ~~~~~~~
  G(\xi) = 1 - \xi^2 + \nu \xi^3 \, .
\eea
It is assumed that $0 < \nu < 2/(3\sqrt{3})$ which guarantees that
$G(\xi)$ has three distinct real roots, $\xi_2$, $\xi_3$, and
$\xi_4$. The roots can be ordered as
$-1 < \xi_2 < 0 < 1 < \xi_3 < \xi_4 < \frac{1}{\nu}$. 

Analyzing the metric \refeq{rotBR}, one finds that in order to keep
the signature Lorentzian and to avoid conical singularities,
the coordinate ranges are required to be
\bea \label{xandy}
  \xi_2 \le x \le \xi_3 \, ,
  ~~~~~~~
  y < \xi_2 \, ,
\eea
and the angular coordinates $\psi$ and $\phi$ must have periodicities 
\bea \label{Deltas}
  \Delta \psi = \Delta \phi 
    = \frac{4\pi \sqrt{F(\xi_2)}}{G'(\xi_2)} \, .
\eea
Furthermore $\xi_1 \ge \xi_3$. If $\xi_1 = \xi_3$ the solution
\refeq{rotBR} describes a black hole with $S^3$ horizon topology. A
coordinate transformation \refRef{Emparan:2001wn} identifies it as the 
Myers-Perry rotating black hole \refRef{Myers:un} in five dimensions
with one rotation parameter set to zero.
In the following we assume $\xi_1 > \xi_3$. Since
the orbit of $g_{\phi\phi}$ then vanishes at both $x=\xi_2$ and
$\xi_3$, there are two distinct conditions imposed on 
$\Delta \phi$. These conditions are solved by setting
\bea \label{xi1}
  \xi_1 = \frac{\xi_4^2 - \xi_2\xi_3}{2\xi_4 - \xi_2 - \xi_3} \, ,
\eea
implying that $\xi_3 < \xi_1 <\xi_4$. Equation \refeq{xi1} can be
viewed as the tuning of the angular momentum to uphold the ring.

The limit $x, y \to \xi_2$ is asymptotic infinity and it can be shown
that the solution is asymptotically flat:
rescale $\psi$ and $\phi$ by taking 
$\tilde{\psi} = 2\pi\psi/\Delta\psi$ and
$\tilde{\phi} = 2\pi\phi/\Delta\phi$, so that 
$\Delta\tilde{\phi} = \Delta\tilde{\psi} = 2\pi$,
and define $\tilde{A} = A \sqrt{G'(\xi_2)}/(2F(\xi_2))$. Via the 
coordinate transformation 
\bea \label{asympcoords}
  \zeta = \frac{\sqrt{\xi_2-y}}{\tilde{A}(x-y)}
  ~~~~~\rom{and}~~~~~
  \eta  = \frac{\sqrt{x-\xi_2}}{\tilde{A}(x-y)} \, ,
\eea
the asymptotic metric can be written as
\bea \label{asympmetric}
  ds^2_\rom{asymp} = - dt^2 + d\zeta^2 + \zeta^2 d\tilde{\psi}^2
     + d\eta^2 + \eta^2 d\tilde{\phi}^2 \, .
\eea

The Killing vector $\p/\p t$ vanishes when $y \to - \infty$, and since
the metric is regular here the coordinate $Y= -1/y$ can naturally be
extended past $Y=0$ allowing $Y$ to take negative values. The
coordinates break down at $y = \xi_4$,  where $g_{yy}$ blows
up. However, this is just a coordinate singularity that can be removed
by a change of coordinates. In fact, one finds that there is an
$S^2 \times S^1$ horizon at $y = \xi_4$ and behind it an $S^1$
curvature singularity is hiding at $y=\xi_1$. The region $y > \xi_4$
is the ergoregion. The ergosurface at $Y=0$ is regular and
has topology $S^2 \times S^1$.

Locally, the rotating black ring is
expected to look like a boosted black string, and indeed the near
singularity behavior of the boosted black string matches that of the
black ring (up to numerical factors and distortion, see also
sections \ref{s:physquan}-\ref{s:boosted}).  

The physical quantities such as the ADM mass, the angular momentum,
and the surface gravity are given for the black ring in
\refRef{Emparan:2001wn}. 
Dimensionless quantities can be formed by multiplying the physical
quantities by suitable powers of the mass. For the angular momentum
$J$ and the horizon area $\mathcal{A}$ we have
\bea \label{JM}
  \frac{J^2}{M^3} 
  &=& 
  \frac{32}{27 \pi} 
  \frac{(\xi_4-\xi_2)^3}
  {(\xi_3-\xi_2)(2\xi_4-\xi_2-\xi_3)^2} \, , \\[3mm]
  \label{AM}
  \frac{\mathcal{A}^2}{M^3} 
  &=& 
  \frac{2048 \pi}{27} 
  \frac{(\xi_3-\xi_2)(\xi_4-\xi_3)}{(2\xi_4-\xi_2-\xi_3)^2} \, .
\eea
We have used \refeq{xi1} to eliminate $\xi_1$ from these expressions,
and we set $G=c=1$ throughout the paper.
In \refsect{s:chargedBR}, we compute the physical quantities
for the charged black ring, and the mass and angular momentum for the
neutral black ring can then be obtained by taking the transformation
parameter $\beta$ to zero (see eqs.~\refeq{themass} and \refeq{theangm}).  


\subsection{An Exact Result for the Black Ring}
\label{s:exact}
\begin{figure}[t!]
  \begin{center}
    \includegraphics[width=11.cm]{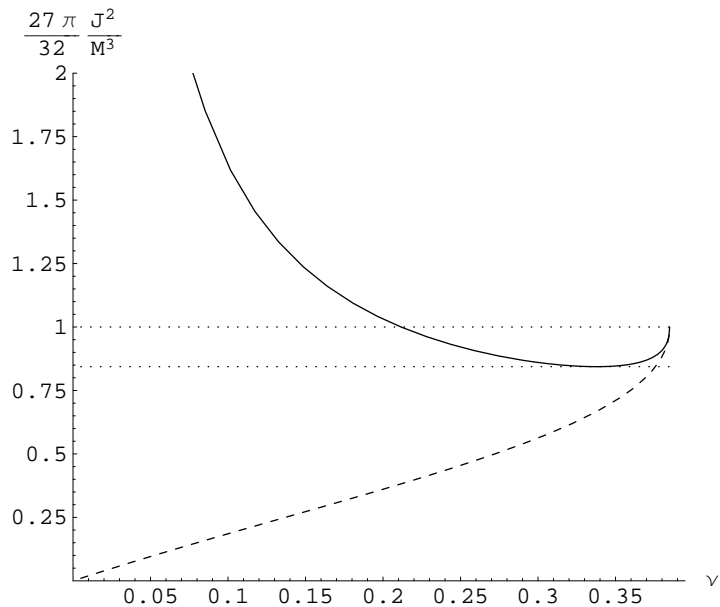}
  \end{center}
  \caption{As in Ref.~\refRef{Emparan:2001wn}, we plot the
    dimensionless ratio 
    $\frac{27 \pi}{32}\,\frac{J^2}{M^3}$ versus $\nu$ for 
    $0 < \nu < 2/(3\sqrt{3})$. The solid line is the ratio for the
    black ring and the dashed line is the ratio for a Myers-Perry
    black hole with only one nonzero angular momentum. The dotted
    lines represent the constant functions $1$ and $27/32$. For the
    black ring, $\frac{27 \pi}{32}\,\frac{J^2}{M^3} = 1$  
    at $\nu \approx 0.211645$.}
  \label{angmom}
\end{figure}

Plotting the dimensionless ratio of angular momentum to mass
as function of $\nu$, Emparan and Reall found that $J^2/M^3$ has a
global minimum, 
\beastar
  \frac{J^2}{M^3} \ge k\, \frac{32}{27 \pi} \, ,
\eeastar
and they estimated $k \approx 0.8437$ \refRef{Emparan:2001wn}. 
We reproduce the plot in \reffig{angmom}.
In this section we determine the exact value of $k$ to be $k=27/32$.
The black ring with $J^2/M^3 = 1/\pi$ is the solution that
maximizes the dimensionless measure of entropy,
$\mathcal{A}/M^{3/2}$. 

The quantities $J^2/M^3$ and $\mathcal{A}^2/M^3$ are given in
\refeq{JM} and \refeq{AM} in terms of the roots, $\xi_2$, $\xi_3$, and
$\xi_4$, of the cubic equation  
\bea \label{cubic}
  \nu \xi^3 - \xi^2 + 1 = 0 \, .
\eea
We assume that $0 < \nu < 2/(3\sqrt{3})$ in order for the 
equation to have three distinct real roots. 
For $\nu = 2/(3\sqrt{3})$, the roots $\xi_3$ and $\xi_4$
coincide. 
Using standard methods for obtaining the roots of a cubic equation, we  
find
\beastar
  \xi_2 &=& \frac{1}{3\nu} 
            \left(
              1 
              - \cos{\textstyle{\frac{\theta}{3}}} 
              - \sqrt{3} \sin{\textstyle{\frac{\theta}{3}}}              
            \right) \, , \\
  \xi_3 &=& \frac{1}{3\nu} 
            \left(
              1 
              - \cos{\textstyle{\frac{\theta}{3}}} 
              + \sqrt{3} \sin\textstyle{\frac{\theta}{3}}              
            \right) \, , \\
  \xi_4 &=& \frac{1}{3\nu} 
            \left(
              1 
              + 2 \cos\textstyle{\frac{\theta}{3}} 
            \right) \, ,
\eeastar
where
\bea \label{theta}
  \theta =  \cos^{-1}\left( 1- \frac{27}{2} \nu^2 \right) \, .
\eea
We can now write the dimensionless ratio of angular momentum to mass
in terms of $\theta$ as 
\bea \label{ratio}
 \frac{27 \pi}{32}\,\frac{J^2}{M^3} 
 = \frac{1}{24}
   \left[ 
     8 + 6\sqrt{3}\csc\left(\textstyle{\frac{2\theta}{3}}\right) 
       + \sec^2\left(\textstyle{\frac{\theta}{3}}\right) 
   \right] \, .
\eea
The global minimum of $\frac{27 \pi}{32}\,\frac{J^2}{M^3}$ for the
black ring (see \reffig{angmom}) is found by extremizing the right
hand side of \refeq{ratio} with respect to $\theta$, and one finds that
$\theta = 3 \cos^{-1}{(2/\sqrt{7})}$ corresponds to the minimum in
\reffig{angmom}.
Solving for $\nu$ using \refeq{theta} we get
\bea \label{nu0}
  \nu_0 = 
  \frac{1}{3} \sqrt{\frac{2}{3} + \frac{20}{21\sqrt{7}}}
  \approx
  0.3377 \, .
\eea
Evaluating \refeq{ratio} at $\nu = \nu_0$ we find the minimum
value $k$ to be
\beastar
  k \equiv
  \frac{27 \pi}{32}\,\frac{J^2}{M^3}\Bigg|_{\nu_0}
  = \frac{27}{32} = 0.84375 \, ,
\eeastar
giving the simple result
\bea \label{MJ}
  M^3 &\le &\pi\, J^2
\eea
for the black ring.
It is peculiar that the value $k = 27/32$ --- a number produced by
extremizing a function which depends solely of the roots of the cubic
equation \refeq{cubic} --- cancels exactly the factor $32/27$ which
comes from the normalization of the mass and the angular momentum. It
would be interesting to understand if there is any significance to
this cancellation. 

In terms of $\theta$, the dimensionless ratio of the horizon area and
the mass is  
\beastar
  \frac{\mathcal{A}^2}{M^3} 
  = \frac{1024 \pi}{81} 
    \tan\left(\textstyle{\frac{\theta}{3}}\right)
    \left[ 
      \sqrt{3}
      -\tan\left(\textstyle{\frac{\theta}{3}}\right) 
    \right] \, .
\eeastar
In the given range, this function has a global maximum for  
$\nu = \nu_0$ given in \refeq{nu0}, and we find
\beastar
  \frac{\mathcal{A}^2}{M^3}\Bigg|_{\nu_0} 
  = 
  \frac{256 \pi}{27} \, .
\eeastar
In conclusion, the black ring with $\nu = \nu_0$ is the black
ring with minimum angular momentum and maximum entropy for the mass
given. 
As $J$ increases, $\mathcal{A}$ decreases, for fixed mass (see also
Fig.~3 of \refRef{Emparan:2001wn}). 

We find that the value of $\nu$ for which 
$\frac{27 \pi}{32}\,\frac{J^2}{M^3} = 1$ is approximately 
$\nu \approx 0.211645$ (see \reffig{angmom}).\footnote{
The value for $\nu$ given in \refRef{Emparan:2001wn} had a minor
typo.}   
One finds that for 
$\frac{1}{\pi} \le \frac{J^2}{M^3} \le \frac{32}{27 \pi}$
there are three distinct solutions with the same asymptotic values $M$
and $J$:
one is the Myers-Perry spherical black hole with $J_\phi = 0$, the two
others are black ring solutions differentiated, for instance, by their 
entropies. 


\setcounter{equation}{0}
\section{Review of Hassan-Sen Transformations}
\label{s:review}

In the early 90s it was shown 
\refRef{Veneziano:1991ek,Meissner:1991zj,%
Sen:1991zi,Sen:1991cn,Gasperini:1991qy}
that in any string theory the space of classical solutions that are
independent of $d$ of the spacetime coordinates has an
\mbox{O($d-1,1$)$\times$O($d-1,1$)} symmetry (or O($d$)$\times$O($d$)
symmetry if the $d$ dimensions are all spatial), where the first
factor acts on the left-movers and the second factor acts on the
right-movers. 

Hassan and Sen \refRef{Hassan:1991mq} showed that in heterotic
string theory the group of transformations can be extended so that the
group acting on the right-movers includes a subset of the 16 internal
coordinates. If the signature of the $d$ coordinates is Lorentzian
and the background gauge fields are neutral under $p$ of the U(1)
generators of the gauge group, the group of transformations is  
O($d-1,1$)$\otimes$O($d+p-1,1$). 
These transformations can be used to generate new
inequivalent classical solutions from known classical solutions.

The symmetry can be realized explicitly for the low energy effective
action, but is valid to all orders in $\al'$. Hassan and Sen
\refRef{Hassan:1991mq} applied the transformations to a magnetic
6-brane solution in ten dimensions to generate a new solution of
heterotic supergravity carrying independent electric and magnetic
charges as well as antisymmetric tensor field charge. Also, starting
from the neutral Kerr black hole in four dimensions, Sen
found a charged rotating black hole solution with
nontrivial dilaton, magnetic fields, and antisymmetric tensor
field \refRef{Sen:1992ua}. Many other solutions have been generated using these
transformations.  

We shall be interested in classical solutions in $D$ spacetime 
dimensions, hence $10-D$ of the ten dimensions for the heterotic
string have been compactified; massless excitations from the
compactification and higher derivative terms are not included in the
effective action, which is given by 
\bea \label{HetS}
  S = \int d^Dx \sqrt{-G} \, e^{-\Phi}
      \Big( R^{(D)}  
      + \nabla_\mu\Phi\nabla^\mu\Phi 
      - \frac{1}{12}H_{\mu\nu\rho}H^{\mu\nu\rho}
      - \frac{1}{8} F_{\mu\nu} F^{\mu\nu}
      \Big) \, .
\eea
We consider only U(1) gauge fields, and just a single U(1)
component of the gauge fields has been included in the action
\refeq{HetS}. The antisymmetric 3-form field $H$ includes the U(1)
Chern-Simons term, 
\bea \label{theH}
  H_{\mu\nu\rho} = (\p_\mu B_{\nu\rho} + \rom{cyclic~permutations}) 
    -\frac{1}{4}(A_\mu F_{\nu\rho} + \rom{cyclic~permutations}) \, .
\eea
Throughout the metric $G_{\mu\nu}$ refers to the string frame
metric. In $D$ dimensions the Einstein metric is related to the 
string metric by 
$G^\rom{E}_{\mu\nu} = e^{-\frac{2}{D-2}\Phi}G_{\mu\nu}$.

We apply the Hassan-Sen transformations to classical solutions that
are independent of the time-direction $x^0$ and (at least) one spatial 
direction $x^1$. The transformations of interest to us involve only
the $(x^0,x^1)$-part of the metric, which we denote by 
$\hat{G}_{ab}$, $a,b=0,1$, the 01-part of the antisymmetric tensor
$\hat{B}_{ab}$ and the gauge fields $\hat{A}_{a}$.
Given such a solution $(G_{\mu\nu},\Phi,A_\mu,B_{\mu\nu})$, where $G$
and $B$ are block diagonal, ie.\ $G_{ai} = B_{ai} = 0$ for all 
$i \ne 0,1$, the transformed solution
$(\bar{G}_{\mu\nu},\bar{\Phi},\bar{A}_\mu,\bar{B}_{\mu\nu})$ 
is computed as follows. Define a $2 \times 2$ matrix $K$ as
\bea \label{K}
  K_{ab} = - \hat{B}_{ab} - \hat{G}_{ab} 
           - \frac{1}{4}\hat{A}_a \hat{A}_b
\eea
and a $5 \times 5$ matrix $M$,
\bea \label{M}
  M = \left(
      \begin{array}{ccccc}
        (K^T - \eta)\hat{G}^{-1}(K - \eta)
        && (K^T - \eta)\hat{G}^{-1}(K + \eta)
	&& -(K^T-\eta)\hat{G}^{-1}\hat{A} \\[1mm]
	(K^T + \eta)\hat{G}^{-1}(K - \eta)
	&& (K^T + \eta)\hat{G}^{-1}(K + \eta)
	&& -(K^T+\eta)\hat{G}^{-1}\hat{A} \\[1mm]
        -\hat{A}^T\hat{G}^{-1}(K-\eta)
	&& -\hat{A}^T\hat{G}^{-1}(K+\eta)
	&& \hat{A}^T\hat{G}^{-1}\hat{A}
      \end{array}
      \right)
\eea
where $\eta = \rom{diag}(-1,1)$. In addition to the above assumptions
we assume for convenience that $\hat{A}=0$ for the
original solution. We have $d=2$ and $p=1$ and the symmetry on
the space of solutions is O(1,1)$\otimes$O(2,1). Writing  
$\Omega \in \rom{O(1,1)} \otimes \rom{O(2,1)}$ as
\beastar
  \Omega = \left(
           \begin{array}{cc}
	   S & 0 \\
           0 & R
           \end{array} 
           \right) \, ,
\eeastar
we can choose $R$ to be on the form 
\bea \label{R}
  R = \left(
      \begin{array}{ccc}
      \cosh{\al_2} & \sinh{\al_2} & 0 \\
      \sinh{\al_2} & \cosh{\al_2} & 0 \\
      0            & 0            & 1 
      \end{array} 
      \right)
      \left(
      \begin{array}{ccc}
      \cosh{\al_1} & 0 & \sinh{\al_1} \\
      0            & 1 & 0            \\
      \sinh{\al_1} & 0 & \cosh{\al_1}  
      \end{array} 
      \right) \, ,
\eea
where $\al_1$ parametrizes boosts that mixes the $0$-direction
with the internal coordinate, and $\al_2$ parametrizes boosts in
$01$-space. The O(1,1)-transformations are Lorentz boosts of
the solution in the 01-plane and we choose $S$ to be the identity
matrix.

The Hassan-Sen transformation acts on the solution to give 
\beastar
  \bar{M} & = & \Omega M \Omega^T \\
  \bar{\Phi} & = & \Phi + \ln{\sqrt{\det{\bar{G}}/\det{G}}}  
\eeastar
The $(x^0,x^1)$-components of the new metric and the fields,
$\bar{A}_a$ and $\bar{B}_{01}$, are given
implicitly by $\bar{M}$ and can be extracted using \refeq{K} and
\refeq{M} with $G$, $K$, and $A$ replaced by $\bar{G}$, $\bar{K}$, and 
$\bar{A}$. All other field components are unchanged by the
transformation. 
In \refappe{s:static} and \refappe{s:stationary} we give the explicit
transformed solution in terms of the original solutions.


\subsubsection*{Remark: Hassan-Sen transformations with
$\al_1=0$} 
Let $(G,\Phi,B=0)$ be a static solution satisfying the
above assumptions. 
It is well-known \refRef{Horne:1991cn} that when
applying a Lorentz boost with parameter $\al$   
($dt \to dt \cosh{\al} + dz \sinh{\al}$ and
$dz \to dt \sinh{\al} + dz \cosh{\al}$)
and then T-dualizing in the $1$-direction, one obtains 
a new solution $(G',\Phi',B')$ where the linear
momentum created by the boost is exchanged for an
$H$-charge.\footnote{The fields are required to fall off appropriately
at infinity.} 
Boosting the solution $(G',\Phi',B')$ in $01$-space
with the same parameter $\al$ gives a new solution $(G'',\Phi'',B'')$
and then T-dualizing again, we find that the 
resulting solution is exactly the Hassan-Sen transformed solution 
$(\bar{G},\bar{\Phi},\bar{B})$ with $\al_1=0$ and $\al_2 = 2\al$. 
This also holds true if the original metric has $G_{01} \ne 0$. 

If $G_{11}=1$ identically for the original static solution
$(G,\Phi,B=0)$ then the last T-duality transformation has no
effect: for $G_{11}=1$ the solution 
$(G'',\Phi'',B'') = (\bar{G},\bar{\Phi},\bar{B})_{\al_1=0}$
is invariant under T-duality. 
In \refsect{s:charBS} we give an example of a self-T-dual 
charged black string. 


\section{The Charged Black Ring}
\label{s:chargedBR}

\subsection{The Solution}
\label{s:soln}

The rotating black ring solution \refeq{rotBR} has three Killing
vectors corresponding to the coordinates $t$, $\psi$, and $\phi$.
We apply the Hassan-Sen transformation with $\beta = \al_1/2$ and
$\al_2 = 0$ to the $(t,\psi)$-part of the black ring solution
\refeq{rotBR} to find a solution of the theory \refeq{HetS} with
$D=5$. The transformation is given explicitly in
\refappe{s:stationary}. The transformed solution is  
\bea 
  \nonumber
  ds^2 &=& -\frac{F(x)}{F(y)\,h(x,y)^2} 
    \left( 
      dt 
      + \sqrt{\frac{\nu}{\xi_1}} \, 
        \frac{\xi_2-y}{A} \cosh^2{\beta} \, d\psi \,
    \right)^2 \\[2mm] \label{twBR1}
    &&\hspace{1.4cm}
    + \frac{1}{A^2 (x-y)^2}
    \bigg[ 
      - F(x) 
      \left( 
      G(y) \, d\psi^2
      + \frac{F(y)}{G(y)}\, dy^2
      \right) \\[2mm] \nonumber
      &&\hspace{4.2cm}
      + F(y)^2
      \left( 
      \frac{dx^2}{G(x)} 
      + \frac{G(x)}{F(x)}\, d\phi^2
      \right)
      \bigg] \, ,
\eea
with fields
\bea \label{twBR2}
  \begin{array}{rclcrcl}
  e^{-\Phi} &=& h(x,y) \, ,
  &~~~&
  \displaystyle B_{t\psi} &\displaystyle = & 
  \displaystyle \sinh^2{\beta}\,\sqrt{\frac{\nu}{\xi_1}}\,
           \frac{F(x)\,(\xi_2 - y)}{A \, F(y) \, h(x,y)} \, , \\[7mm]
  A_t &=&  \displaystyle 
           \frac{(x - y)\,\sinh{2\beta}}{\xi_1 F(y) \, h(x,y)}\, ,
  &~~~&
  A_\psi &=& \displaystyle -\sinh{2\beta}\,\sqrt{\frac{\nu}{\xi_1}}\,
           \frac{F(x)\,(\xi_2 - y)}{A \, F(y) \, h(x,y)} \, ,
  \end{array}
\eea
The functions $F$ and $G$ are given in \refeq{FandG} and  
\bea \label{h}
  h(x,y) = 1 + \frac{x-y}{\xi_1 F(y)} \,\sinh^2{\beta} \, .
\eea
The analysis of the metric with respect to signature and regularity
works out exactly as for the neutral case.
The coordinates $x$ and $y$ are restricted to the regions
\refeq{xandy} and the coordinates $\psi$ and $\phi$ are periodic with
the periods given in \refeq{Deltas}. 
We note that for $x$ and $y$ in the coordinate regions \refeq{xandy},
the function $h$ in equation \refeq{h} is strictly positive.

The asymptotic region is at $x,y \to \xi_2$. Since $h \to 1$ at
infinity, the coordinate transformation \refeq{asympcoords} takes the 
asymptotic metric to the form \refeq{asympmetric} after the
appropriate rescalings. Thus the transformed metric is asymptotically 
flat. 

For $\xi_1 = \xi_3$, the coordinate transformation given in
\refRef{Emparan:2001wn} takes the solution given by \refeq{twBR1}
and \refeq{twBR2} to the solution obtained by applying the Hassan-Sen 
transformation of \refsect{s:stationary} to the five dimensional
Myers-Perry black hole with only one nonzero rotation parameter. This
solution is the five dimensional analog of the charged rotating black
hole in four dimensions found by Sen \refRef{Sen:1992ua}. It can be
generalized to a charged solution with two independent angular momenta
by applying a Hassan-Sen transformation to the general five
dimensional Myers-Perry rotating black hole. 

In the following we assume that $\xi_1 > \xi_3$. Just as in the
case of the neutral black ring, regularity requires $\xi_1$ to be
given by \refeq{xi1}.

The transformed solution is regular at $y \to -\infty$,
so defining $Y=-1/y$ we can extend the coordinate region to include $Y<0$
just as for the neutral black ring.
At $y=\xi_4$, the metric component $g_{yy}$ blows up while
the fields stay finite. By a  slight modification of the coordinate
transformation $(t,\psi) \to (v,\chi)$ given in
\refRef{Emparan:2001wn} we obtain new coordinates for which the metric 
is regular at $y=\xi_4$. The new coordinates --- valid for $y>\xi_1$
--- are defined as
\beastar
  d\chi  &=& d\psi    + \frac{\sqrt{-F(y)}}{G(y)} dy  \\
  dt &=& dv + 
        \sqrt{\frac{\nu}{\xi_1}}\,
        \frac{(y-\xi_2)\,\sqrt{-F(y)}}{A\,G(y)}\,
        \cosh^2{\beta}\, dy \, ,
\eeastar
so the $t,\psi,y$-part of the metric becomes
\beastar
  ds^2_{v\chi y} &=& 
   -\frac{F(x)}{F(y)\,h(x,y)^2} 
    \left( 
      dv 
      + \sqrt{\frac{\nu}{\xi_1}} \, 
        \frac{\xi_2-y}{A} \cosh^2{\beta} \, d\chi \,
    \right)^2 \\[4mm]
    && \hspace{2.7cm}
    + \frac{F(x)}{A^2 (x-y)^2}
    \bigg[ 
      - G(y) d\chi^2
      + 2\sqrt{-F(y)} \, d\chi dy
    \bigg] \, .
\eeastar
The Killing vector $(\p/\p t)$ vanishes at $Y = 0$, so the region 
$y > \xi_4$ is the ergoregion. The determinant 
$g_{tt}g_{\psi\psi} - g_{t\psi}^2$ has a zero at $y = \xi_4$ and since
we know that the metric is regular here, the constant-$(v,y)$ surface
at $y = \xi_4$ defines the event horizon. There is no inner
horizon. Both the ergosurface (defined as the constant-$(t,Y)$ surface
at $Y = 0$) and the horizon are topologically $S^2 \times S^1$.  
The curvature blows up at $y = \xi_1$, and the dilaton is singular
there; this corresponds to a spacelike $S^1$ curvature singularity in
the metric. 


\subsection{Physical Properties}
\label{s:physquan}

Going to the Einstein metric,
$G^\rom{E}_{\mu\nu} = e^{-\frac{2}{3}\Phi} G_{\mu\nu}$,
and using the next to leading order behavior of the asymptotic metric
we compute the ADM mass and the angular momentum
\refRef{Myers:un}
\bea \label{themass}
  M &=& \frac{3\pi}{2 A^2}\, 
  \frac{(1+\frac{4}{3} \sinh^2{\beta})\,(\xi_1 - \xi_2)}
  {\xi_1^2\, \nu\, (\xi_3 - \xi_2)\,(\xi_4 - \xi_2)} \, ,
  \\[3mm] \label{theangm}
  J &=& \frac{2\pi}{A^3} \,
  \frac{(\xi_1 - \xi_2)^{5/2}\,\cosh^2{\beta}}
  {\xi_1^3\, \nu^{3/2}\, (\xi_3 - \xi_2)^2\,(\xi_4 - \xi_2)^2} \, ,
\eea
which reduce to the values for the neutral ring for $\beta \to 0$.
Also, the black ring has an asymptotic electric U(1) charge given by 
(charges are normalized as in \refRef{Horne:1991cn})
\beastar
  Q =   \frac{1}{16 \pi}
	\int_{S^3~\rom{at}~\infty} 
        \hspace{-7mm} e^{-\Phi} \star F
        =
	\frac{\pi\,(\xi_1-\xi_2)\,\sinh{2\beta}}
        {A^2\, \xi_1^2\, \nu\, (\xi_3-\xi_2)\,(\xi_4-\xi_2)} \, .
\eeastar
The dimensionless ratio of angular momentum and mass is given by
\bea \label{angmass}
  \frac{J^2}{M^3} &=& 
  \frac{\cosh^4{\beta}}{(1+\frac{4}{3} \sinh^2{\beta})^3}\,
  \left[\frac{J^2}{M^3}\right]_\rom{\beta=0}
  \ge \frac{1}{\pi}\,
    \frac{\cosh^4{\beta}}{(1+\frac{4}{3} \sinh^2{\beta})^3}\, ,
\eea
where we have used the result \refeq{MJ} for lower bound on $J^2/M^3$
for the neutral black ring, and the dimensionless ratio of charge to
mass is  
\bea \label{charmass}
  \frac{Q}{M} &=& 
  \frac{2 \sinh{2\beta}}{3 (1+\frac{4}{3} \sinh^2{\beta})} \, .
\eea
We note that by taking $\beta$ large we can make $J^2/M^3$
arbitrarily small while $Q/M$ approaches a constant. Thus
the charge helps holding up the black ring allowing us to make the
angular momentum arbitrarily small. This was of course not possible
for the neutral black ring.

Surprisingly, the ratio $Q/M$ is independent of $\nu$. In
fact, we notice that the right hand side of \refeq{charmass} is always
less than 1, so that for all $\beta$ we have
\bea \label{QandM}
  |Q| \le M 
\eea
with equality in the limit $\beta \to \infty$. Contrary to other
solutions with angular momentum and charge, this bound does not
involve the angular momentum.

As a one dimensional object in a five dimensional asymptotically flat
spacetime, the black ring can carry local --- but not global ---
fundamental string charge associated with the 3-form field $H$.
Using \refeq{theH} we find that $H$ has only one nonzero component,
\beastar
  H_{t\psi y} = - \frac{\sqrt{\frac{\nu}{\xi_1}}\,
                    (\xi_1 - x)^2\,\sinh^2{\beta}}
                  {A\,\big[\xi_1+x\sinh^2{\beta}-y\cosh^2{\beta}\big]^2}
  \, ,
\eeastar
and it gives rise to the local fundamental string charge
\beastar
  q_H &=& \frac{1}{16\pi}\int e^{-\Phi}\star H =
      \frac{\sinh^2{\beta}\,(\xi_1-\xi_2)^{1/2}}
            {4\, A\, \xi_1\, \nu^{1/2}\, (\xi_4-\xi_2) } \, ,
\eeastar
where the integral is over a two sphere parametrized by
$x$ and $\phi$ at a constant $\psi$-cut around of the ring.
In the limit $\beta \to \infty$, the dimensionless ratio $q_H^2/M$
diverges. 

From the leading order behavior of the field 
$A_{\tilde{\psi}} = A_\psi\,\Delta \psi/(2\pi)$ at infinity we
find the magnetic moment $\mu$ of the black ring. In spherical
coordinates with radial coordinate $\rho$ and a polar coordinate
$\theta$ we have for large $\rho$ 
\beastar
  A_{\tilde{\psi}} &=& -\frac{\bar{\mu}\,\sin^2{\theta}}{\rho^2} 
\eeastar
In analog to the normalization of the charges, we normalize the
magnetic moment as $\mu = \bar{\mu}\,A_3/(16\pi)$ where $A_3$ is
the area of a unit three sphere.
We find
\beastar
  \mu &=& \frac{\pi\,(\xi_1-\xi_2)^{5/2}\,\sinh{2\beta}}
            {A^3\,\xi_1^3\,\nu^{3/2}(\xi_3-\xi_2)^2(\xi_4-\xi_2)^2}
  \, .
\eeastar
The $\nu$ dependence cancels in $\mu/J$ so that the ratio
depends only on $\beta$.
The gyromagnetic ratio $g$ is defined as $g = 2 \mu M/(Q J)$ and we
find
\bea \label{gyro}
  g &=& \frac{3 (1+\frac{4}{3} \sinh^2{\beta})}{2 \cosh^2{\beta}} 
  \, ,
\eea
so that $g$ is independent of $\nu$.
We see from \refeq{gyro} that for the charged black ring, the
$g$-factor can take values between $3/2$ and $2$. 
The same bounds have been found on the gyromagnetic
ratio for the string theory solution describing a dilatonic rotating
charged black hole in a four dimensional asymptotically flat
spacetime \refRef{Horne:zy}. It should however be noted that there
is an ambiguity in the normalization of the magnetic moment; changing
the normalization of $\mu$ changes the $g$-factor.

The area of the event horizon is 
\beastar
  \mathcal{A} &=& \frac{16 \pi^2}{A^3}
    \frac{(\xi_1 - \xi_2)\,(\xi_4 - \xi_1)^{3/2}\,\cosh^2\beta}
         {\nu^{3/2}\,\xi_1^3\,(\xi_3 - \xi_2)\,
          (\xi_4 - \xi_2)^2\,(\xi_4 - \xi_3)} \, .
\eeastar
As a function of $\beta$, the dimensionless ratio
$\mathcal{A}/M^{3/2}$ is maximized for  
$\beta = 0$: for a given mass the neutral black ring always has higher 
entropy than the charged black ring, and
increasing the charge while keeping the mass fixed,
the horizon area decreases. This is qualitatively the same
behavior as for a charged spherical black hole.

Associated with the horizon is a Killing vector field
\bea \label{killing}
  \frac{\p}{\p v} \,+\, 
        \frac{A\,\xi_1^{1/2}}
             {\nu^{1/2}\,(\xi_4-\xi_2)\,\cosh^2{\beta}}
  \;\frac{\p}{\p \chi} \, .
\eea
Outside the horizon where the original coordinates are valid, the
Killing vector field is given by
\beastar
  \frac{\p}{\p t} \,+\, 
        \frac{A\,\nu^{1/2} \xi_1 (\xi_3-\xi_2)}
             {2\,(\xi_1-\xi_2)^{1/2}\,\cosh^2{\beta}}
  \;\frac{\p}{\p \tilde\psi} \, .
\eeastar
Using the Killing field \refeq{killing} we compute the surface gravity
\bea \label{BRsurfgrav} 
  \kappa &=& \frac{A\,(\xi_4-\xi_3)\,\nu^{1/2}\,\xi_1}
             {2\,(\xi_4-\xi_1)^{1/2}\,\cosh^2{\beta}} \, .
\eea

Near the singularity, $y = \xi_1 + \ep$ for small $\ep > 0$, the
metric takes the form 
\beastar
  ds^2 \sim 
    +\ep (dv - d\chi)^2 + d\chi^2 + \sqrt{\ep}\, d\chi d\ep 
    + \ep^2 d{\Omega'}_2^2 \, ,
\eeastar
where we have ignored numerical constants and $x$-dependence
(for example, the $(x,\phi)$-part of the metric is only
topologically a two sphere). For the fields we find near the
singularity 
\beastar
  e^{-\Phi} \sim O(\ep^{-1})
  ~~~~\rom{and}~~~~
  B_{tz}, A_t, A_z \sim O(1)
\eeastar
We shall compare this behavior with the near singularity behavior of 
the Hassan-Sen transformed boosted black string (see
\refsect{s:boosted}). 


\subsection{Extremal Limit}
\label{s:extring}

In the limit $\nu \to 0$ the $\xi_i$'s behave as
\beastar
  \xi_1 \sim \frac{1}{2\nu} + \frac{\nu}{4} \, ,
  ~~~~~
  \xi_2 \sim -1 + \frac{\nu}{2} \, ,
  ~~~~~
  \xi_3 \sim 1 + \frac{\nu}{2} \, ,
  ~~~~~
  \xi_4 \sim \frac{1}{\nu} - \nu \, ,
  ~~~~~
\eeastar
The ratio $J^2/M^3$ approaches zero when $\beta  \to \infty$, but it
diverges for $\nu \to 0$ (see \reffig{angmom}).
We find an extremal limit of the charged black ring by taking the
limit $\beta \to \infty$ keeping $\lambda \equiv \nu\, e^{2\beta}$ fixed.
The extremal metric is
\beastar
  ds^2 &=& -\frac{1}{\big[1+\frac{\lambda}{2}(x-y)\big]^2} 
       \left(dt - \frac{\lambda (1+y)}{2\sqrt{2} A} \, d\psi \right)^2 \\
       & & \hspace{2cm}
       + \frac{1}{A^2 (x-y)^2}
       \Big[ (y^2-1)\, d\psi^2 + \frac{dy^2}{y^2-1} 
             + \frac{dx^2}{1-x^2} + (1-x^2) \, d\phi^2 \Big] 
\eeastar
and the fields are
\beastar
  \begin{array}{rclcrcl}
  e^{-\Phi} &=& 1 + \frac{\lambda}{2}(x-y) \, ,
  &~~~~&
  B_{t\psi} &=& \displaystyle -\frac{\lambda (1+y)} 
          {2\sqrt{2} A \big[ 1 + \frac{\lambda}{2}(x-y)\big]} \, , \\[7mm]
  A_t &=& \displaystyle \frac{\lambda (x-y)}
          {\big[ 1 + \frac{\lambda}{2}(x-y)\big]}  \, ,
  &~~~~&
  A_\psi &=& \displaystyle \frac{\lambda (1+y)} 
          {\sqrt{2} A \big[ 1 + \frac{\lambda}{2}(x-y)\big]}  \, .
  \end{array}
\eeastar
We now have $-1<x<1$ and $y<-1$, and the periodicities are 
$\Delta \psi = \Delta \phi = 2\pi$. The solution is asymptotically
flat. 
The curvature blows up at $y \to -\infty$ and this is a null
singularity coinciding with the horizon.

The physical quantities for the extremal solution are
\beastar
  \begin{array}{rclrclrcl}
  M &=& \displaystyle \frac{\pi \lambda}{2 A^2} \, , &
  ~~~~~~
  J &=&  \displaystyle\frac{\pi \lambda}{2^{5/2} A^3} \, , &
  ~~~~~~
  Q &=&  \displaystyle \frac{\pi \lambda}{2 A^2} \, , \\[5mm]
  q_H &=&  \displaystyle \frac{\lambda}{8\sqrt{2}\, A} \, , &
  ~~~~~~
  \mu &=&  \displaystyle \frac{\pi \lambda}{2^{5/2} A^3} \, , &
  ~~~~~~
  g &=& 2 \, . 
  \end{array}
\eeastar
Note that the inequality \refeq{QandM} is saturated in the extremal
limit so that $|Q| = M$. Also, for the extremal ring $\mu = J$ and the
$g$-factor is 2. By taking $\lambda$ large we can make $J^2/M^3$
arbitrarily small. 
The horizon area shrinks to zero, however taking the limit of the 
surface gravity \refeq{BRsurfgrav} gives a constant, 
$\kappa \to \frac{\sqrt{2} A}{\lambda}$. This is similar to the
behavior found in \refRefs{Horne:zy,Garfinkle:qj} for slowly rotating
and non-rotating dilatonic charged spherical black hole solutions in
string theory.

Defining $r=-1/y$ and considering small $r>0$ we find that the near
horizon/singularity behavior is 
\beastar
  ds^2 \sim -\frac{4 r^2}{\lambda^2} 
         \left( dt + \frac{\lambda}{2\sqrt{2} A r}\, d\psi \right)^2  
         +\frac{1}{A^2} \left( d\psi^2 + dr^2 + r^2 d\Omega_2^2\right)
\eeastar
with
\beastar
  e^{-\Phi} \sim \frac{\lambda}{2r} \, ,
  ~~~~~
  A_t , A_z , B_{tz} \to \rom{constant} \, .
\eeastar 
We compare this with the near-singularity behavior of the extremal
limit of the charged boosted black string in \refsect{s:boosted}.


\subsection{A Charged Boosted Black String}
\label{s:boosted}

The local behavior of the neutral rotating black ring is like
that of a boosted black string, hence we expect that the charged black 
ring behaves locally as a boosted black string with similar
charges and fields. We check this by comparing the near singularity
behavior of the charged black ring to that of a charged black string
obtained from the boosted black string by the Hassan-Sen
transformation of \refappe{s:stationary}.

The black string metric in five dimensions is the four dimensional
Schwarzschild solution times $\R$,  
\bea \label{neutralBS}
  ds^2 &=& -\Big(1-\frac{r_0}{r}\Big) dt^2 + dz^2 
      + \Big(1-\frac{r_0}{r}\Big)^{-1} dr^2 + r^2 d\Omega_2^2 \, .
\eea
Applying a Lorentz boost to the solution \refeq{neutralBS} by taking  
$dt \to dt \cosh{\al} + dz \sinh{\al}$ and
$dz \to dt \sinh{\al} + dz \cosh{\al}$, we obtain the metric for the
boosted black string
\bea \label{boostedBS}
  \nonumber
  ds^2 &=& -\left( 1-\frac{r_0 \cosh^2{\al}}{r}\right)\, dt^2
           + \frac{r_0 \sinh{2\al}}{r}\, dtdz
           +\left( 1+\frac{r_0 \sinh^2{\al}}{r}\right)\, dz^2 
   \\[2mm]
       & & \hspace{3.5cm}
           +\left( 1-\frac{r_0}{r}\right)^{-1} dr^2 
           +r^2 d\Omega_2^2 \, .
\eea
T-dualizing the metric \refeq{boostedBS} gives the solution for the
non-extremal fundamental black string (see \refsect{s:charBS}). Now
instead apply the Hassan-Sen transformation of \refsect{s:stationary}
to \refeq{boostedBS}. The transformed solution is given by 
\beastar
  ds^2 &=& 
    - \frac{r\,(r-r_0 \cosh^2{\al})}
    {\big[ r+r_0\, \cosh^2{\al}\,\sinh^2{\beta}\big]^2}
    \left(dt - 
      \frac{r_0\, \sinh{2\al} \cosh^2{\beta}}
           {2\big( r-r_0 \cosh^2{\al} \big)}\, dz
    \right)^2 
    \\[2mm]
    & &\hspace{2.5cm}
    +\, \frac{r-r_0}{r-r_0 \,\cosh^2{\al}} \, dz^2
    +\left( 1-\frac{r_0}{r}\right)^{-1} dr^2 
           +r^2 d\Omega_2^2 \, ,
    \\[6mm]
    &&
    \begin{array}{rclcrcl}
      e^{-\Phi} &=& \displaystyle 
      1 + \frac{r_0\, \cosh^2{\al}\,\sinh^2{\beta}}{r} \, ,
      &~~~~~&
      \displaystyle 
      B_{tz} &=& \displaystyle
        -\frac{r_0\, \sinh{2\al}\, \sinh^2{\beta}}
              {2\big[ r+r_0\,\cosh^2{\al}\,\sinh^2{\beta}\big]} \, ,
      \\[6mm]
      A_t &=& \displaystyle
         \frac{r_0\, \cosh^2{\al}\, \sinh{2\beta}}
              {\big[ r+r_0\, \cosh^2{\al}\,\sinh^2{\beta}\big]} \, ,
      &~~~~~&
      \displaystyle
      A_z &=& \displaystyle
         \frac{r_0\, \sinh{2\al}\, \sinh{2\beta}}
              {2\big[ r+r_0\, \cosh^2{\al}\,\sinh^2{\beta}\big]} \, .
    \end{array}
\eeastar
The solution is regular at $r=r_0\,\cosh^2{\al}$ and the coordinate
singularity at $r=r_0$ can be removed by a coordinate transformation
(valid for $r<r_0\,\cosh^2{\al}$)
\beastar
  dv &=& dt 
         + \frac{\sqrt{r_0\,\cosh^2{\al}/r - 1}}{1-r_0/r} \,
           \frac{r_0\, \sinh{2\al}\, \cosh^2{\beta}}
                {2\big( r-r_0 \,\cosh^2{\al}\big)}\,dr \\
  dw &=& dz 
         + \frac{\sqrt{r_0\,\cosh^2{\al}/r - 1}}{1-r_0/r} \,dr  \, .
\eeastar
In these coordinates the metric becomes
\beastar
  ds^2 &=& 
    - \frac{r\,(r-r_0 \cosh^2{\al})}
    {\big[ r+r_0\, \cosh^2{\al}\,\sinh^2{\beta}\big]^2}
    \left(dv - 
      \frac{r_0\, \sinh{2\al} \cosh^2{\beta}}
           {2\big( r-r_0 \,\cosh^2{\al} \big)}\, dw
    \right)^2 
    \\[2mm]
    & &\hspace{2.2cm}
    +\, \frac{r-r_0}{r-r_0 \,\cosh^2{\al}} \, dw^2
    +2 \big(r_0\,\cosh^2{\al}/r - 1\big)^{-1/2}\, drdw
           +r^2 d\Omega_2^2 \, ,
\eeastar
and it is regular at $r=r_0$. 
The Killing vector $\p/\p t$ becomes null at $r=r_0\,\cosh^2{\al}$,
and the determinant $g_{tt}g_{zz} - g_{tz}^2$ vanishes at $r=r_0$;
the region $r_0<r<r_0\,\cosh^2{\al}$ is the ergoregion for
the string and $r=r_0$ defines the horizon. 
There is a curvature singularity at $r=0$.

The physical quantities can be calculated as in the previous section. 
We find (the linear momentum is computed following
\refRef{Horne:1991cn})  
\beastar
  \begin{array}{rclcrcl}
    m &=& \displaystyle
    \frac{r_0}{4}\big[ 1+ \cosh^2{\al}\,\cosh{2\beta}\big] \, ,
    &~~~~~&
    P_z &=& \displaystyle
    \frac{r_0}{8} \sinh{2\al}\,\cosh^2{\beta} \, , \\[4mm]
    q &=& \displaystyle
    \frac{r_0}{4} \cosh^2{\al}\,\sinh{2\beta} \, ,
    &~~~~~&
    Q_H &=& \displaystyle
    -\frac{r_0}{8} \sinh{2\al}\,\sinh^2{\beta}
  \end{array}
\eeastar
and the surface gravity is
\beastar
  \kappa &=& \frac{1}{2r_0\,\cosh{\al}\cosh^2{\beta}} \, .
\eeastar
The near singularity behavior of the solution is (ignoring numerical
constants) 
\beastar
  ds^2 &\sim& + r(dv + dw)^2
      + dw^2 - \sqrt{r}\, drdw +r^2 d\Omega_2^2 \, , \\[3mm]
  && e^{-\Phi} \sim O(r^{-1}) \, ,
  ~~~~~
  B_{tz} , A_t, A_z \sim \rom{constants}
\eeastar
Qualitatively, this agrees with the near singularity behavior of the
charged black ring. For the black ring, the two sphere is distorted
and so is the $S^1$ around the ring.


\subsubsection*{Extremal limit}
There is an extremal limit defined by taking $\beta\to\infty$ while
keeping $\lambda_0 \equiv \frac{r_0}{4} e^{2\beta}$ fixed. The extremal
solution is 
\beastar
  ds^2 &=& 
    - \frac{1}
    {\big[ 1+\lambda_0\,\cosh^2{\al}/r\big]^2}
    \left(dt - 
      \frac{\lambda_0\,\sinh{2\al}} {2 r}dz
    \right)^2 
    + dz^2 + dr^2 +r^2 d\Omega_2^2 \, ,
    \\[6mm]
    &&
    \begin{array}{rclcrcl}
      e^{-\Phi} &=& \displaystyle 
      1 + \frac{\lambda_0\,\cosh^2{\al}}{r} \, ,
      &~~~~~&
      \displaystyle 
      B_{tz} &=& \displaystyle
        -\frac{\lambda_0\,\sinh{2\al}}
              {2\big[ r+\lambda_0\,\cosh^2{\al}\big]} \, ,
      \\[6mm]
      A_t &=& \displaystyle
         \frac{2\lambda_0\, \cosh^2{\al}}
              {\big[ r+\lambda_0\, \cosh^2{\al}\big]} \, ,
      &~~~~~&
      \displaystyle
      A_z &=& \displaystyle
         \frac{\lambda_0\, \sinh{2\al}}
              {\big[ r + \lambda_0\, \cosh^2{\al}\big]} \, .
    \end{array}
\eeastar
At $r=0$ there is a null singularity coinciding with the horizon. The
near singularity behavior for the extremal boosted string is 
\beastar
  ds^2 &=& - \frac{r^2}{\lambda_0^2 \cosh^4{\al}}
           \left(dt - 
             \frac{\lambda_0\,\sinh{2\al}} {2 r}dz
           \right)^2
         + dz^2 + dr^2 +r^2 d\Omega_2^2 \, ,
    \\[5mm]
    &&
      e^{-\Phi} = \frac{\lambda_0\,\cosh^2{\al}}{r} \, ,
      ~~~~~~
      B_{tz}, A_t, A_z \sim \rom{constants}
\eeastar
As expected, this agrees with the near singularity behavior of the
extremal charged black ring.


\section{Black Strings with Charge}
\label{s:charBS}
The most general five dimensional black string solution labelled by
mass, angular momentum, electric charge per length, and fundamental
string charge was analyzed by Mahapatra in
\refRef{Mahapatra:1993gx}. The solution was found by applying a
Hassan-Sen transformation with general $\al_1$ and $\al_2$ to the
rotating black string vacuum solution obtained as the four dimensional
Kerr solution times a flat direction. In this section we consider a
special case where the Hassan-Sen transformation is applied to the
neutral non-rotating black string and we find that in one extremal
limit this solution describes traveling waves in a fundamental string
background. 

Applying the Hassan-Sen transformation of \refsect{s:static} to the
neutral black string \refeq{neutralBS}, we find a solution with the
metric  
\bea \label{HchFchBS1} 
  \hspace{-9mm}
  ds^2 = -\frac{4r(r-r_0) - r_0^2 b^2}{4(r+\tilde{r})^2} \, dt^2
    + \frac{r_0 b}{r+\tilde{r}} \, dtdz 
    + dz^2 
    + \Big(1-\frac{r_0}{r}\Big)^{-1} \, dr^2 
    + r^2 \, d\Omega_2^2 \, , 
\eea
and with fields 
\bea \label{HchFchBS2} 
  e^{-\Phi} = 1+\frac{\tilde{r}}{r} \, ,
  ~~~~~
  B_{tz} = \frac{r_0 b}{2(r+\tilde{r})} \, , 
  ~~~~~
  A_t = \frac{r_0 a}{r+\tilde{r}} \, ,
  ~~~~~
  A_z = 0 \, .
\eea
We have defined
\beastar
  a = \sinh{\al_1} \, , 
  ~~~~~
  b = \cosh{\al_1} \sinh{\al_2} \, ,
  ~~~~~
  \tilde{r} = \frac{r_0}{2}
      \left( \cosh{\al_1} \cosh{\al_2} - 1\right) \, .
\eeastar
There is an $S^2 \times \R$ horizon at  $r=r_0$ and the metric 
\refeq{HchFchBS1} has a curvature singularity at $r=0$.
The fields are regular everywhere, except at $r=0$ where the dilaton
blows up. 
The solution represents a charged black string with mass per length  
$m$, electric charge per length $q$, fundamental string charge $Q_H$,
and linear momentum $P_z$ along the string, given by 
\beastar
  m=(r_0/4) (1+\cosh{\al_1}\cosh{\al_2}) \, ,
  ~~~~
  q = r_0 a/4 \, ,
  ~~~~
  Q_H =  P_z = r_0 b/8 \, . 
\eeastar
Expressed in terms of the three independent physical parameters $m$,
$q$, and $Q_H$, we have  
\beastar
  r_0 = \frac{2}{m}\Big( m^2 - q^2 - 4 Q_H^2 \Big)
\eeastar
so we must require
\bea \label{mqQrel1}
  m^2 \ge q^2 + 4Q_H^2 
\eea
for the solution not to be naked singular.
There are two extremal limits that saturate the inequality
\refeq{mqQrel1}; we study these at the end of this section.

\mlrum
Consider now the solution for a black string with fundamental string
charge in five dimensions
\refRef{Horowitz:1992jp,Horne:1991cn,Horowitz:cd} 
given by
\bea \label{HchBS}
  \nonumber
  ds^2 &=& -\frac{r-r'_0}{r+r'_0 \sinh^2{\al'}} \, dt^2 
      + \frac{r}{r+r'_0 \sinh^2{\al'}} dz^2
      + \Big(1-\frac{r_0'}{r}\Big)^{-1} \, dr^2 
      + r^2 \, d\Omega_2^2 \, \\[1mm] 
  e^{-\Phi} &=& 1 + \frac{r'_0 \sinh^2{\al'}}{r} \, , \\[1mm]\nonumber
  B_{tz} &=& \frac{r'_0 \sinh{2\al'}}{2(r+r'_0 \sinh^2{\al'})} \, ,
\eea
and $A_t = A_z = 0$. 
In the extremal limit $\al' \to \infty$ with $r_0' e^{2\al'}$
constant, this solution describes the fields outside a straight
fundamental string \refRef{Dabholkar:yf,Horowitz:cd}.

The solution \refeq{HchBS} can be obtained from the neutral
black string \refeq{neutralBS} by first applying a Lorentz boost and
then T-dualizing to exchange the linear momentum with $H$-charge
\refRefs{Horne:1991cn,Horowitz:cd,Horowitz:1992jp}.  

Applying a Lorentz boost with boost parameter $\al'$ to the solution
\refeq{HchBS} we obtain a solution describing a boosted black 
string with $H$-charge. As discussed in \refsect{s:review} this is
exactly the same solution as the transformed solution in
\refeqs{HchFchBS1}{HchFchBS2} with $\al_1=0$, $\al_2 = 2\al'$, and
$r_0 = r'_0$, and it is invariant under T-duality. 

\mlrum
Returning to the solution given in \refeq{HchFchBS1} and
\refeq{HchFchBS2} we find that the inequality \refeq{mqQrel1} can be
saturated by taking $\al_1 \to \infty$ with $\al_2$ fixed or vice
versa. We consider both cases in the following.


\subsubsection*{Extremal limit I}
The inequality \refeq{mqQrel1} is saturated by taking 
$\al_2 \to \infty$ while keeping 
$\al_1$ and $R \equiv \frac{1}{4}r_0 e^{\al_2}$ constant. 
In this limit the electric field vanishes, and the $B$-field and the
dilaton behave just as the fields for a straight fundamental
string \refRefs{Horowitz:1992jp,Horne:1991cn,Horowitz:cd}, but whereas
the fundamental string is boost invariant along the $z$-direction,
this solution is not.  

Introducing null coordinates 
$u = \frac{1}{\sqrt{2}}(z+t)$ and $v = \frac{1}{\sqrt{2}}(z-t)$ 
the Einstein metric for this extremal solution can be written
\beastar
  ds^2_\rom{Einstein} &=&
   \left( 1 + \frac{M}{r} \right)^{-1/3}
   \left(2dudv + \frac{2M}{r}du^2 \right) 
   + \left( 1 + \frac{M}{r} \right)^{2/3}  
     \left(dr^2 + r^2 d\Omega_2^2 \right)
\eeastar
where $M = R \cosh{\al_1}$. This matches exactly the traveling wave
solutions found by Garfinkle \refRef{Garfinkle:zj}. Thus the solution
describes constant spherically symmetric traveling waves along a
fundamental black string. 


\subsubsection*{Extremal limit II}
Taking $\al_1 \to \infty$ while keeping $\al_2$
and  $R' \equiv \frac{1}{4}r_0 e^{\al_1}$ constant gives another
extremal solution with charges that saturate the inequality
\refeq{mqQrel1}. This extremal solution is given by the metric
\beastar
  \hspace{-9mm}
  ds^2 = -\frac{r^2 - K^2}{(r+M)^2} \, dt^2
    + \frac{2K}{r+M} \, dtdz 
    + dz^2 
    + dr^2 + r^2 \, d\Omega_2^2 \, ,
\eeastar
where now $M = R' \cosh{\al_2}$ and $K = R' \sinh{\al_2}$, and the 
fields
\beastar
  e^{-\bar{\Phi}} = 1+\frac{M}{r} \, ,
  ~~~~~
  B_{tz} = \frac{K}{r+M} \, , 
  ~~~~~
  A_t = \frac{2\sqrt{M^2 - K^2}}{r+M} \, ,
  ~~~~~
  A_z = 0 \, .
\eeastar
The physical mass per length is $m = M/2$, the electrical charge per
length is $q = \frac{1}{2}\sqrt{M^2-K^2}$, and the fundamental string
charge $Q_h = K/4$ equals the linear momentum $P_z$ along the string.


\setcounter{equation}{0}
\section{Discussion}
\label{s:disc}

We have constructed a charged black ring solution to five dimensional
heterotic supergravity and studied its properties. The solution is
labelled by mass, angular momentum, and electric charge, and it also
carries local fundamental string charge. 
We found that the lower bound on the dimensionless ratio $J^2/M^3$
could be arbitrarily small, because the charge would help supporting
the ring from collapsing.

As mentioned in \refsect{s:soln}, the solution given
by \refeq{twBR1} and \refeq{twBR2} describes a charged rotating black
ring for $\xi_1 > \xi_3$ and a charged rotating black hole for 
$\xi_1 = \xi_3$. The latter can also be obtained by applying the
Hassan-Sen transformation of \refappe{s:stationary} to the five
dimensional Myers-Perry black hole with one of the rotation parameters 
set to zero. For a given $\beta$, the ratio $|Q|/M$ is fixed, and
one can then use the parameter $\nu$ to set the value
of $J^2/M^3$. This holds for both the charged black ring and the
charged spherical black hole solutions. 
The behavior of $J^2/M^3$ as a function of $\nu$ is as for the neutral
black ring (see \reffig{angmom}), just rescaled by a $\beta$-dependent
factor. Hence there exists a range of values of $J^2/M^3$ such that
there are three distinct solutions with the same asymptotic charges:
one is the Hassan-Sen transform of the Myers-Perry black hole with
just one nonzero angular momentum, and the two others are black ring
solutions (one has greater horizon area than the other). This is an
extension of the non-uniqueness result of Ref.~\refRef{Emparan:2001wn}
to charged solutions of low energy heterotic string theory.

The charged black hole found here can be generalized by applying a
Hassan-Sen transformation to the general five dimensional Myers-Perry
black hole with two independent rotation parameters.

For the charged black ring we found a curious relationship between
the mass and the charge: $|Q| \le M$. The inequality is saturated in
the extremal limit, encouraging that the extremal solution may
actually be supersymmetric. This requires further investigation.

The stability of the black ring solutions is an open question. The
black rings resemble thin black strings, in particular for small
$\nu$, so one may expect to find a classical Gregory-Laflamme
instability mode \refRef{Gregory:vy}.
If it exists, such an instability mode would generate ripples in the
horizon, and due to the rotation the ripples would radiate away and
one would expect the ring to collapse to a rotating black hole. 

\mlrum
\mlrum
Very recently, Hong and Teo \refRef{Hong:2003gx}
presented a new form of the C-metric, and they expressed the neutral
black ring metric \refeq{rotBR} in terms of these new
coordinates. Some expressions in this paper simplify when the
coordinates of Hong and Teo are used; in particular, the calculation
of the exact value of the lower bound of $J^2/M^3$ simplifies
drastically. 


\section*{Acknowledgement}

It is a great pleasure to thank Gary Horowitz for innumerable helpful
discussions and guidance throughout this work, and for comments on the
draft. I am grateful to Roberto Emparan and Harvey Reall for comments
and also for bringing Ref.~\refRef{Hong:2003gx} to my attention. I
would like to thank the Niels Bohr Institute for hospitality during
the early stages of the project. This work was supported by the Danish
Research Agency and NSF grant PHY-0070895.


\appendix

\setcounter{equation}{0}
\section{Hassan-Sen Transformed Solutions}

In this appendix we give explicit expressions for solutions obtained
by the Hassan-Sen solution generating technique reviewed in
\refsect{s:review}. 

\subsection{Transformation of a Static Metric}
\label{s:static}
Let $(G_{\mu\nu},\Phi)$ be a static solution 
independent of the time-direction $x^0$ and the spatial direction
$x^1$ with no gauge fields and no antisymmetric tensor
fields. The transformations considered here involve only 
the $(x^0,x^1)$-part of the metric. We assume that the metric is block
diagonal, ie.\ $G_{ai} = 0$ for all $i \ne 0,1$ and $a,b=0,1$.
The transformed solution is then given by the metric 
\beastar
  \bar{G}_{00} &=& \frac{4}{X(G)} 
    \Big\{
      \cosh^2{\al_1}\,\sinh^2{\al_2} \; G_{00}^{-1} \big(1+G_{00}\big)^2
      + G_{11}^{-1} \big[(1+G_{11}) + \cosh{\al_2}(1-G_{11})\big]^2
    \Big\} \, ,\\[5mm]
\bar{G}_{01} &=& \frac{4\sinh{\al_2}}{X(G)} 
    \Big\{
      \cosh{\al_1}\, G_{00}^{-1} \big(1+G_{00}\big)
        \big[(1-G_{00})+\cosh{\al_1}\cosh{\al_2}(1+G_{00})\big] \\
    & & \hspace{5.3cm}
      + \, G_{11}^{-1} \big(1-G_{11}\big)
        \big[(1+G_{11}) + \cosh{\al_2}(1-G_{11})\big]
    \Big\} \, ,  \\[5mm]
\bar{G}_{11} &=& \frac{4}{X(G)} 
    \Big\{
      G_{00}^{-1}\,
      \big[(1-G_{00})+\cosh{\al_1}\cosh{\al_2}(1+G_{00})\big]^2
      + \sinh^2{\al_2}\ G_{11}^{-1} \big(1-G_{11}\big)^2
    \Big\} \, ,
\eeastar
where
\beastar
  X(G) &=& G_{00}^{-1}G_{11}^{-1}
    \Big\{
    (1-G_{00})\big[(1+G_{11}) + \cosh{\al_2}(1-G_{11})\big] \\
    & & \hspace{3.8cm}
    + \cosh{\al_1} (1+G_{00})
     \big[(1-G_{11}) + \cosh{\al_2}(1+G_{11})\big]
    \Big\}^2 \, ;
\eeastar
the dilaton
\beastar
  \bar{\Phi} &=& \Phi + \ln{\sqrt{\frac{16}{G_{00}G_{11}X(G)}}} \, ;  
\eeastar
the gauge fields
\beastar
  \bar{A}_0 &=& \frac{2 \sinh{\al_1} (1+G_{00}) 
    \big[(1+G_{11}) + \cosh{\al_2}(1-G_{11})\big]}
   {\sqrt{G_{00}G_{11}X(G)}} \, , \\
  \bar{A}_1 &=& \frac{2 \sinh{\al_1}\sinh{\al_2} (1+G_{00})(1-G_{11})}
   {\sqrt{G_{00}G_{11}X(G)}}  \, ;
\eeastar
and an antisymmetric tensor field
\beastar
  \bar{B}_{01} = \frac{\sinh{\al_2}}{\sqrt{G_{00}G_{11}X(G)}}
    \Big[ (1-G_{00})(1-G_{11}) + \cosh{\al_1}(1+G_{00})(1+G_{11})
    \Big] \, .
\eeastar
Even when the original solution is asymptotically flat, the
transformed metric is not necessarily so. 
However, if the original solution is asymptotically flat such
that $G_{00} \to -1$ and $G_{11} \to 1$ at infinity, then 
$\bar{G}_{00} \to -1$, $\bar{G}_{01} \to 0$,
and $\bar{G}_{11} \to 1$ at infinity, so the transformed solution is
also asymptotically flat. 

If $G_{11} = 1$ identically, 
then $\bar{G}_{11} = 1$ identically and there will be no magnetic U(1)
field, $\bar{A}_1 = 0$. 
Note that U(1) gauge fields are generated only if $\al_1 \ne 0$ and
that an antisymmetric tensor field is generated only if 
$\al_2 \ne 0$. Also, $\bar{G}_{01}$ vanishes for $\al_2 = 0$.


\subsection{Transformation of a Stationary Metric 
($\al_2 = 0$)} 
\label{s:stationary}
Let $(G_{\mu\nu},\Phi)$ be a solution satisfying the
conditions from \refsect{s:static} only that now we allow for an
off-diagonal term $G_{01} \ne 0$. 
Setting $\beta = \al_1/2$ and $\al_2 = 0$, the transformed solution is 
\beastar
  \begin{array}{rclcrcl}
  \bar{G}_{00} &=& \displaystyle
    \frac{G_{00}}{\big[ 1 + (1+G_{00})\sinh^2{\beta}\big]^2} \\[6mm]
  \bar{G}_{01} &=& \displaystyle
    \frac{G_{01} \cosh^2{\beta}}
         {\big[ 1+ (1+G_{00})\sinh^2{\beta}\big]^2} \\[6mm]
  \bar{G}_{11} &=& \displaystyle
    \frac{G_{00}^{-1} \, G_{01}^2 \cosh^4{\beta}}
         {\big[ 1+ (1+G_{00})\sinh^2{\beta}\big]^2} 
    + G_{00}^{-1} \, \det{\hat{G}} \, , \\[6mm]
  \bar{\Phi} &=& \displaystyle
    \Phi - \ln{\big[ 1 + (1+G_{00})\sinh^2{\beta}\big]} \, , 
  &~~&
  \bar{B}_{01} & = &   \displaystyle
                 -\frac{G_{01}\,\sinh^2{\beta}}
                 {1 + (1+G_{00})\sinh^2{\beta}} \\[6mm]
  \bar{A}_0 & = & \displaystyle
            \frac{(1+G_{00})\,\sinh{2\beta}}
                 {1 + (1+G_{00})\sinh^2{\beta}} \, ,
  &~~&
  \bar{A}_1 & = & \displaystyle
            \frac{G_{01}\,\sinh{2\beta}}
                 {1 + (1+G_{00})\sinh^2{\beta}} \, .
  \end{array}
\eeastar
For $G_{01} = 0$, this is just the $\al_2 = 0$ case of the previous
section.

If the original solution is asymptotically flat with
$G_{00} \to -1$ and $G_{01} \to 0$ at infinity, then the new solution
is asymptotically flat, and the matter fields approach zero at
infinity. 


\end{document}